\documentclass[aps,prl,twocolumn,superscriptaddress,showpacs,preprintnumbers,amsmath,amssymb,floatfix]{revtex4}

\usepackage{graphicx}
\usepackage{dcolumn}
\usepackage{bm}
\usepackage{url} 
\usepackage{amsmath} 
\usepackage{amssymb}

\renewcommand\Re{\operatorname{\mathfrak{Re}}}
\renewcommand\Im{\operatorname{\mathfrak{Im}}}

\begin{document}


\title{\quad\\[1.0cm] Effect of nuclear interactions of neutral kaons
  on $CP$ asymmetry measurements }

\affiliation{Korea University, Seoul}
\affiliation{Faculty of Mathematics and Physics, 
University of Ljubljana, Ljubljana}
\affiliation{J. Stefan Institute, Ljubljana}
\affiliation{Institute for Theoretical and Experimental Physics, Moscow}

\author{B.~R.~Ko}\affiliation{Korea University, Seoul}
\author{E.~Won}\email[Corresponding author. ]{eunil@hep.korea.ac.kr}
\affiliation{Korea University, Seoul} 
\author{B.~Golob}\affiliation{Faculty of Mathematics and Physics, University of Ljubljana, Ljubljana}\affiliation{J. Stefan Institute, Ljubljana} 
\author{P.~Pakhlov}\affiliation{Institute for Theoretical and Experimental Physics, Moscow} 

\begin{abstract}
We examine the effect of the difference in nuclear interactions 
of ${K}^0$ and
$\bar{K}^0$ mesons on the measurement of 
$CP$ asymmetry for
experiments at $e^+e^-$ colliders - 
charm and $B$-meson factories. We
find that this effect on $CP$ asymmetry can be as large as 0.3\%,
and therefore sufficiently significant in interpreting measurements of $CP$
asymmetry when neutral kaons are present in the final state.

\end{abstract}


\maketitle

{\renewcommand{\thefootnote}{\fnsymbol{footnote}}}
\setcounter{footnote}{0}

Modern high-statistics $B$ factories discovered the joint violation of
charge-conjugation and parity ($CP$) in $B$-meson decay modes. In some $B^0$
decays~\cite{ref:belle_jpsi},
large $CP$ violation induced by $B^0-\bar{B}^0$ mixing is
observed to be consistent with the
predictions of the standard model (SM) and the Kobayashi-Maskawa
ansatz~\cite{ref:KM}.  Smaller, direct $CP$ violations, attributed to 
the interference of different amplitudes, but without mixing
have also been reported~\cite{ref:kpi,ref:direct}. SM
predictions for the direct $CP$ violation in 
many charmed-meson decays
are typically of $\mathcal{O}(10^{-3})$~\cite{ref:SMcharm}.  However,
the present accuracy of measurements of $CP$ asymmetry in
$D$ meson decays is close to their SM expectations. For example, in the decay
$D^+ \to K^0_S\pi^+$~\cite{ref:conj}, the statistical sensitivity on
the measured $CP$ asymmetry (of $\approx$0.2\%)~\cite{ref:acp_ko} is
slightly smaller than the effect expected in the SM
of (0.332$\pm$0.006)\% from $K^0-\bar{K}^0$ mixing~\cite{ref:pdg2010}.
Experiments at future high-luminosity $B$
factories and at the LHC are likely to reach the sensitivity needed to observe
$CP$ violation in some $D$ decay modes.

The measured asymmetries of $B$ or $D$ mesons for decays which has
$K^0_S$ in their final states,
can be mimicked (or diluted) by differences
between 
$K^0$ and $\bar{K}^0$ interactions with detector material. 
The probability of an inelastic
interaction of a neutral kaon in the detector depends on
the strangeness of the kaon at any point along its path,
which is due to 
oscillations in kaon strangeness and different nuclear cross sections
for ${K}^0$ and $\bar{K}^0$. Hence the total efficiency to observe a
final state $K_S^0$ differs from that expected for 
either ${K}^0$ or $\bar{K}^0$.
This effect is related to the coherent regeneration of
neutral kaons~\cite{ref:pais}.  
This kind of contribution may be non-negligible for
precise measurements of direct $CP$ violation in $B$ and $D$ decays,
and  
also important in the determination of $\phi_3$ in the $B^+ \to D^0 K^+
\to (K^0_S \pi^+ \pi^-)_DK^+$ transition~\cite{ref:ggsz} and 
in a precise measurement
of $D^0-\bar{D}^0$ mixing in the $K^0_S \pi^+ \pi^-$ final state, as 
the Dalitz distribution would be distorted by the $K^0$ interaction.

In this paper, we evaluate the effect of the difference in 
nuclear interactions of neutral kaons on measurements of $CP$ asymmetry
performed at charm and $B$ factories, or will be
carried out at the near future high-luminosity $B$ factories.  Our study
represents an extension and more detailed description of the method
used to estimate the effect of ${K}^0/\bar{K}^0$ interactions in
material in Ref.~\cite{ref:acp_ko}. We also note that the detector-simulation 
program {\sc Geant4}~\cite{ref:geant4}, commonly used in
high energy physics experiments, does not take into account the effect
considered in this paper, as the ${K}^0$ and $\bar{K}^0$ are projected
onto the $K^0_S$ or $K^0_L$ components at their production point
rather than at their points of $\pi \pi$ decay. The
time-dependent $K^0-\bar{K}^0$ oscillations are thereby ignored in 
{\sc Geant4}. A similar effect in $D^0-\bar{D}^0$
oscillations was found to be small in the mass and lifetime differences 
between $D^0$ and $\bar{D}^0$~\cite{ref:silva}. The aim of this paper 
is to approximately estimate the magnitude of the effect due to 
the difference in $K^0$ and $\bar{K}^0$
nuclear interactions under conditions of
current and future experiments, and bring this issue to the 
simulation developers for possible inclusion in 
programs such as {\sc Geant4}. The method and result can serve as an 
estimate of systematic uncertainty for measurements neglecting the 
effect, or as a starting point for more refined calculations to 
be used in the future experiments in order to correct for the effect. 

Let us consider production of some meson $\mathcal{P}$
and its antimeson $\bar{\mathcal{P}}$  in $e^+ e^-$
collisions, each followed by its decay into states containing
a neutral kaon, and observed through the $K^0_S \rightarrow \pi^+\pi^-$ 
or $\pi^0 \pi^0$ mode:
\begin{eqnarray}
&& \mathcal{P} \to K^0_S + X,
\nonumber \\
&& \bar{\mathcal{P}} \to {K}^0_S + \bar{X}. 
\nonumber 
\label{eq:setup}
\end{eqnarray}
$\mathcal{P}$ can be a charmed or $B$ meson. 
For certain charmed meson decays, there is a small contribution
from doubly Cabibbo-suppressed decays that we ignore, in 
our main calculation, but assign a systematic uncertainty 
for this assumption. 
The $CP$ asymmetry in the $\mathcal{P}$ decays is defined as
\begin{eqnarray}
A^{\mathcal{P}\rightarrow K^0_S + X}_{CP} = 
\frac
{\int d\Gamma^{\mathcal{P}\rightarrow K^0_S + X} - 
 \int d\Gamma^{\bar{\mathcal{P}} \rightarrow {K}^0_S + \bar{X}}}  
{\int d\Gamma^{\mathcal{P}\rightarrow K^0_S + X} + 
 \int d\Gamma^{\bar{\mathcal{P}}\rightarrow {K}^0_S + \bar{X}}}, 
\label{eq:acp}
\end{eqnarray}
where $\Gamma$ denotes the partial decay width. We  
assume that the
production point is surrounded by a cylindrical structure of material, typically
used in a collider detector environment, such as a beam pipe and several
thin layers of vertex detectors. 

 To obtain the time development of neutral kaons in matter, 
we use the calculation carried out in Refs. \cite{ref:good,ref:fetscher}.
The time evolution of amplitudes in the $K^0_L$ and $K^0_S$ basis,
as given in Ref.~\cite{ref:fetscher}, becomes  
\begin{eqnarray}
\alpha_\textrm{L}(t) &=& \textrm{e}^{-i\Sigma\cdot t}
\Bigg[
\alpha^0_\textrm{L}
\cos{\Big(\frac{\Delta \lambda}{2}\sqrt{1+4r^2}~t \Big)}
\nonumber \\
&-&
i\frac{\alpha^0_\textrm{L}+2r\alpha^0_\textrm{S}}{\sqrt{1+4r^2}} 
\sin{\Big(\frac{\Delta \lambda}{2}\sqrt{1+4r^2}~t \Big)}
\Bigg],
\nonumber \\
\alpha_\textrm{S}( t) &=& \textrm{e}^{-i\Sigma\cdot t}
\Bigg[
\alpha^0_\textrm{S}
\cos{\Big(\frac{\Delta \lambda}{2}\sqrt{1+4r^2}~t \Big)}
\nonumber \\
&+&
i\frac{\alpha^0_\textrm{S}-2r\alpha^0_\textrm{L}}{\sqrt{1+4r^2}} 
\sin{\Big(\frac{\Delta \lambda}{2}\sqrt{1+4r^2}~t \Big)}
\Bigg],
\nonumber \\
{\rm where}
\nonumber \\
\Sigma &\equiv&\frac{1}{2}(\lambda_L + \lambda_S + \chi + \bar{\chi}),
\nonumber \\
\Delta \lambda&=&\lambda_L - \lambda_S 
\nonumber \\
&=&
\Delta m - \frac{i}{2} \Delta \Gamma=
(m_L-m_S) - \frac{i}{2}(\Gamma_L - \Gamma_S),
\nonumber \\
\Delta \chi&=&\chi - \bar{\chi} = 
-\frac{2\pi \mathcal{N}}{m}{\Delta f}=-\frac{2\pi \mathcal{N}}{m}(f-\bar{f}).
\label{eq:good}
\end{eqnarray}
The quantities $\alpha_\textrm{L}(t)$ and 
$\alpha_\textrm{S}(t)$ are the amplitudes for 
finding states as a $K^0_L$ and ${K}^0_S$ at some time $t$, respectively,
and $\alpha^0_\textrm{L}$ and $\alpha^0_\textrm{S}$ are those states at $t$=0, 
where $t$ refers to their proper times.
The masses $m_L$ and $m_S$, 
and decay widths $\Gamma_L$ and $\Gamma_S$ refer to  
$K^0_L$ and $K^0_S$, respectively.
The quantity $m$ in $\Delta \chi$ denotes the mass of the $K^0$ and $\bar{K}^0$. 
The volume density of material is $\mathcal{N}$$\equiv$$\frac{\rho N_A}{M}$,
where $\rho$ is the mass density. 
$N_A$ is Avogadro's number, and $M$ is the mean molar mass.
The quantities $f$ and $\bar{f}$ are the forward scattering amplitudes 
of $K^0$ and $\bar{K}^0$, respectively.
The parameter $r$ is called the regeneration parameter, defined
as $r=\frac{1}{2}\frac{\Delta \chi}{\Delta \lambda}$, and
its magnitude is generally small, typically in the order
of $10^{-2}$.
Expanding $\alpha_\textrm{L}(t)$ and $\alpha_\textrm{S}(t)$ up to the first 
order in $r$, 
we obtain 
\begin{eqnarray}
\alpha_\textrm{L}(t) &=& \xi_L(t) \alpha^0_\textrm{L}  
+ \zeta(t) \alpha^0_\textrm{S}  r,
\nonumber \\
\alpha_\textrm{S}(t) &=& \xi_S(t) \alpha^0_\textrm{S}   
+ \zeta(t) \alpha^0_\textrm{L}  r,
\label{eq:good_linear}
\end{eqnarray}
where 
$\xi_{L,S}(t) = \frac{1}{2}\textrm{e}^{-\frac{i}{2}(\chi + \bar{\chi})t}
\textrm{e}^{-i\lambda_{L,S} t}
$ and
$\zeta(t) = \frac{1}{2}{\textrm{e}^{-\frac{i}{2}(\chi + \bar{\chi})t}}
\Big(
\textrm{e}^{-i\lambda_L t}
-
\textrm{e}^{-i\lambda_S t}
\Big)$. 
From these relations, the amplitudes following the 
passage of
several layers of detector material,
can be obtained iteratively as follows:
\begin{eqnarray}
\alpha_\textrm{L}(t_j) &=& \xi_L(t_j-t_{j-1}) \alpha_\textrm{L}(t_{j-1}) 
\nonumber \\
&+& \zeta(t_j-t_{j-1}) \alpha_\textrm{S}(t_{j-1}) r_j
\nonumber \\
\alpha_\textrm{S}(t_j) &=& \xi_S(t_j-t_{j-1}) \alpha_\textrm{S}(t_{j-1}) 
\nonumber \\
&+&\zeta(t_j-t_{j-1}) \alpha_\textrm{L}(t_{j-1}) r_j
\label{eq:iter}
\end{eqnarray}
where index $j$ refers to the layer of material
last penetrated. This follows because neutral kaons pass through
several layers of
vacuum and material before they decay.
The number of terms in Eq.~(\ref{eq:iter}) increases
rapidly as the number of detector layers increases, and
it is squared when 
$|\alpha_\textrm{L,S}(t)|^2$
are computed to obtain the probability. We evaluate
all terms using
the symbolic calculation program 
{\sc Mathematica}~\cite{ref:mathematica}.
Our dilution effect ($A_\mathcal{D}$) can be extracted from the 
total asymmetry ($A_T$),
which is incorporated in $K^0$ regeneration. Without $CP$ violation 
in $\mathcal{P}\rightarrow K^0_S + X$ decay itself, 
the $A_T$ in the decay can be expressed as
\begin{eqnarray}
A^{\mathcal{P}\rightarrow K^0_S + X}_T &\equiv& 
\frac{
\int R(t)~d\Gamma^{\mathcal{P}\rightarrow K^0_S + X}
-
\int \bar{R}(t)~d\Gamma^{\bar{\mathcal{P}}\rightarrow {K}^0_S + \bar{X}}
}
{
\int R(t)~d\Gamma^{\mathcal{P}\rightarrow K^0_S + X}
+
\int \bar{R}(t)~d\Gamma^{\bar{\mathcal{P}}\rightarrow {K}^0_S + \bar{X}}
}
\nonumber \\
&\cong& A^{K^0}_{CP} + A_{\mathcal{D}} + A_{\textrm{int}},
\label{eq:k0asm}
\end{eqnarray}
where $R(t)$ and $\bar{R}(t)$, the
two-pion decay rates for initial $K^0$ and $\bar{K}^0$, 
respectively, can be expressed as 
$R_S|\alpha_{\rm S}(t)+\eta\alpha_{\rm L}(t)|^2$.
$R_S$ is the time-independent decay rate of the $K^0_S$ eigenstate,
and the ratio of amplitudes 
$\eta = \mathcal{M}(K^0_L \rightarrow \pi^+ \pi^-)/\mathcal{M}(K^0_S \rightarrow \pi^+\pi^-)$.  
The first term in Eq.~(\ref{eq:k0asm}), $A^{K^0}_{CP}$,
is the asymmetry due to $K^0-\bar{K}^0$ mixing which 
is not of primary interest in this paper, and thus can be subtracted.
The third term, $A_{\textrm{int}}$ is 
the asymmetry from interference between the $CP$ violation in $K^0$ mixing 
and the material related
amplitudes, and is  
expected to be of 
$\mathcal{O}(|r \eta|)\approx10^{-5}$. We estimate the third term 
numerically as $\approx10^{-6}$,
and therefore ignore it. 
Hence, $A_T$ reduces to $A_{\mathcal{D}}$ if the $CP$ violation effect due to 
$K^0-\bar{K^0}$ mixing in $A_T$ is removed, thereby setting the 
parameter $\eta$=0.
Approximating $\Re(\Delta f)/\mathfrak{Im}(\Delta f)=1$,
and $\Delta m\approx\frac{1}{2}\Delta \Gamma$, 
$A_{\mathcal{D}}$ can be expressed as
\begin{eqnarray}
A_{\mathcal{D}} \propto - 
\Im(\Delta f) \propto \sigma(\bar{K}^0 N) - \sigma({K}^0 N),
\label{eq:k0asmII}
\end{eqnarray}
where $N$ refers to the 
atomic nucleon of the detector material.
\begin{figure}[htbp]
\includegraphics[width=0.45\textwidth]{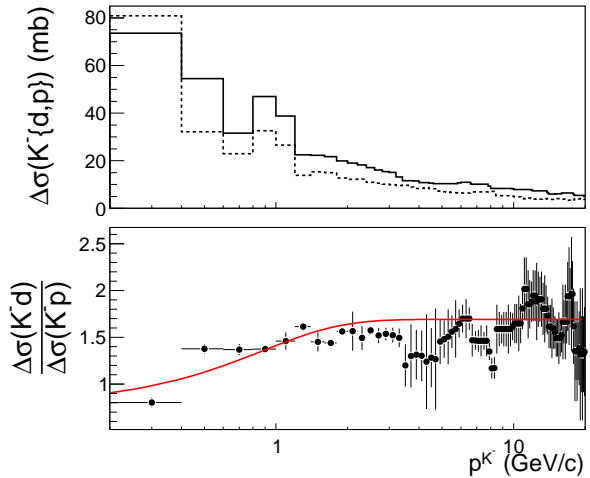}
\caption{
The $\Delta \sigma (K^-\{d,p\})$ 
= $\sigma (K^-\{d,p\}) - \sigma(K^+\{d,p\})$
values as a function of kaon
momentum, obtained from~\cite{ref:pdg2010} for the proton (dotted lines)
and the deuteron data (solid) are shown in the top plot. The ratio of two
cross section differences as a function of kaon momentum is shown in
the bottom plot (solid circles), together with the fit using 
the error function (curve).
}
\label{fig:xsec_pdg}
\end{figure}

To compute results for Eq.~(\ref{eq:k0asmII}) taking into account effects
of nuclear screening~\cite{ref:glauber},
we adopt an empirical scaling
law based on measurements in C, Al, Cu, Sn, and Pb for neutral kaon momenta 
($p^{\bar{K}^{0}}$)
between 20 and 140 GeV/$c$~\cite{ref:gsponer}:
\begin{eqnarray}
\Delta \sigma (\bar{K}^0 N) 
&\equiv& \sigma(\bar{K}^0 N) - \sigma(K^0 N)
\nonumber \\
&=& 
\frac{23.2A^{0.758\pm0.003}}
{[p^{\bar{K}^{0}}(\textrm{GeV}/c)]^{0.614}}~\textrm{mb}, 
\label{eq:a}
\end{eqnarray}
where $A$ is the atomic number and 
0.758 accounts for nuclear screening. 
The scaling of $A^{0.758}$ in Eq.~(\ref{eq:a}) also describes 
Pb, Cu, and C data quite well down to 5 GeV/$c$~\cite{ref:gsponer}.
The deuteron data in Ref.~\cite{ref:caroll} also agree well with 
the prediction of Eq.~(\ref{eq:a})
for $A=2$ from 50 to 200 GeV/$c$. We extend the scaling down to lower
momenta assuming  isospin symmetry of nuclear interactions, 
$\sigma(\bar{K}^0 \textrm{n})$$\cong$$\sigma(K^- \textrm{p})$ and 
$\sigma({K}^0 \textrm{p})$$\cong$$\sigma(K^+ \textrm{n})$.
We approximate $\sigma(K^+ \textrm{n})$$\cong$$\sigma(K^+ \textrm{p})$ 
to improve the estimation of $A_{\mathcal{D}}$, and 
this assumption is consistent with
measurements~\cite{ref:pdg2010}.
(Symbols p and n correspond to the proton and neutron, respectively.)
Using experimental results 
for $\sigma(K^- \textrm{p})$ and $\sigma(K^+ \textrm{p})$ 
from Ref.~\cite{ref:pdg2010}, 
we obtain $\Delta\sigma(K^-\{\textrm{d},\textrm{p}\})$, where d denotes
deuteron,
with  
$\Delta \sigma(K^-d)$$\equiv$$\sigma (K^- d)
- \sigma(K^+ d)$
and
$\Delta \sigma(K^-p)$$\equiv$$\sigma (K^- p)
- \sigma(K^+ p)$.
Figure~\ref{fig:xsec_pdg} 
shows $\Delta\sigma(K^-\{\textrm{d},\textrm{p}\})$ (top) 
and the ratio of the two, 
$\Delta \sigma(K^-d) /
\Delta \sigma(K^-p)$ 
 (bottom), 
as a function of the kaon momentum. We fit the ratio of 
$\Delta\sigma(K^- \textrm{d})$ to $\Delta\sigma(K^- \textrm{p})$ 
using an empirical function while keeping the 
nuclear screening term 
$A^{0.758}$ fixed.
The value of 
$\chi^2/{\rm d.o.f}$ is approximately 2, 
indicating our modeling of the ratio of $\Delta\sigma(K^- \textrm{d})$ to 
$\Delta\sigma(K^- \textrm{p})$ is 
not unreasonable, so that Eq.~(\ref{eq:a}), obtained in the high-momentum 
range, can be scaled down to 1 GeV/$c$ and below. 
Using the fit, Eq.~(\ref{eq:a}) is altered as follows
\begin{eqnarray}
\Delta\sigma(\bar{K}^0 N)
&=& \frac{A^{0.758}\Delta \sigma (K^- \textrm{p})}
	    {1+1.252 \textrm{e}^{-1.841 p^{{K}^{-}} (\textrm{GeV/}c)}}~\textrm{mb}, 
\label{eq:brko}
\end{eqnarray}
where $p^{{K}^{-}}$ is the momentum of $K^-$.
We use Eq.~(\ref{eq:brko}) in the
numerical calculation of Eq.~(\ref{eq:k0asmII}). 
The numerator in Eq.~(\ref{eq:brko}) should
extrapolate the screening effect to atoms in the detector material
we use in Table~\ref{table:configuration}, and the denominator
reflects the low-momentum behavior of the difference in cross section
between the proton and deuteron data. We compared
our scaling method with the experimental data~\cite{ref:cplear}, and
found a good agreement.

\begin{figure}[htbp]
\mbox{
\includegraphics[width=1.0\textwidth,width=0.5\textwidth]{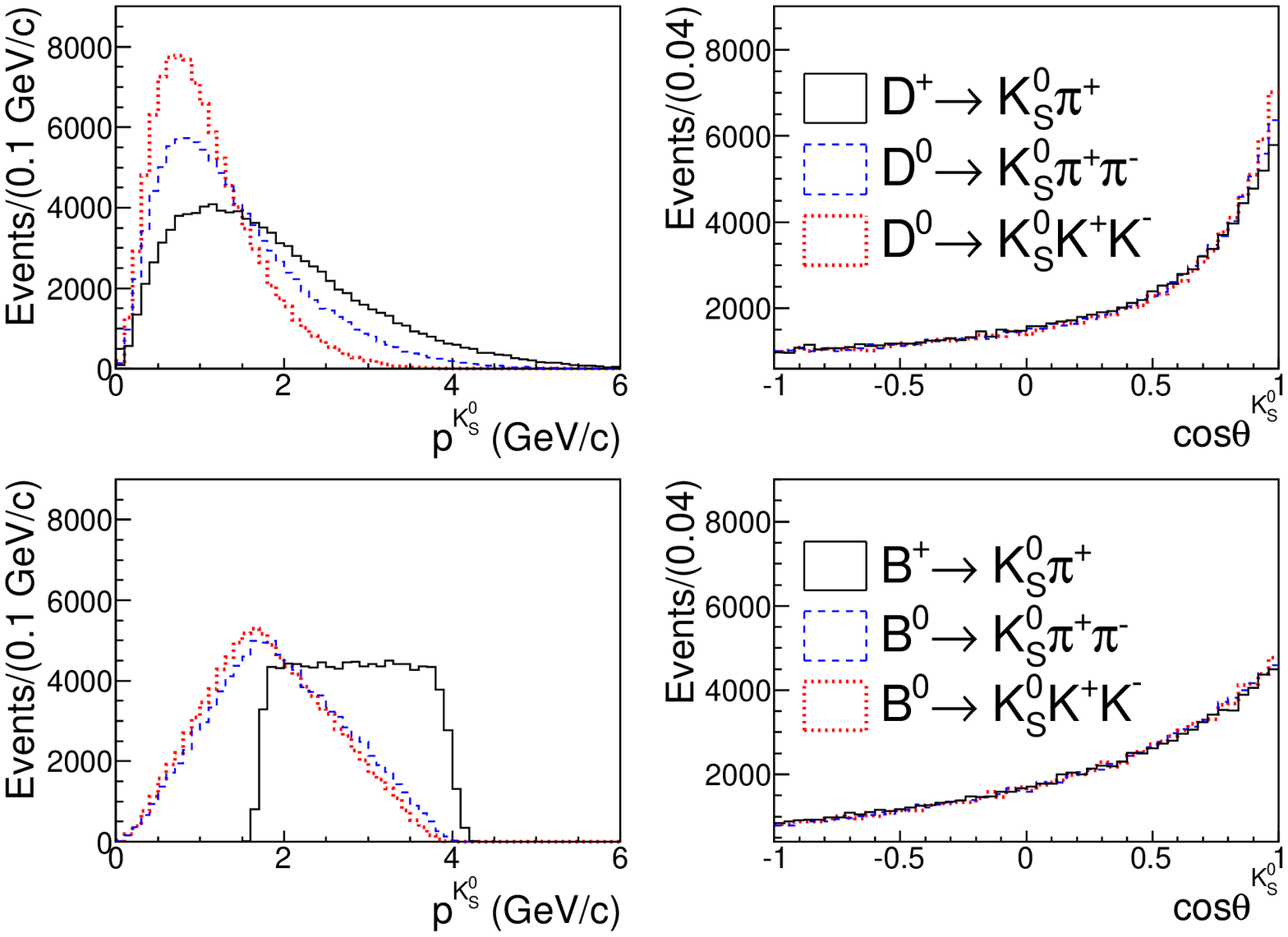}
}
\mbox{
\includegraphics[width=1.0\textwidth,width=0.5\textwidth]{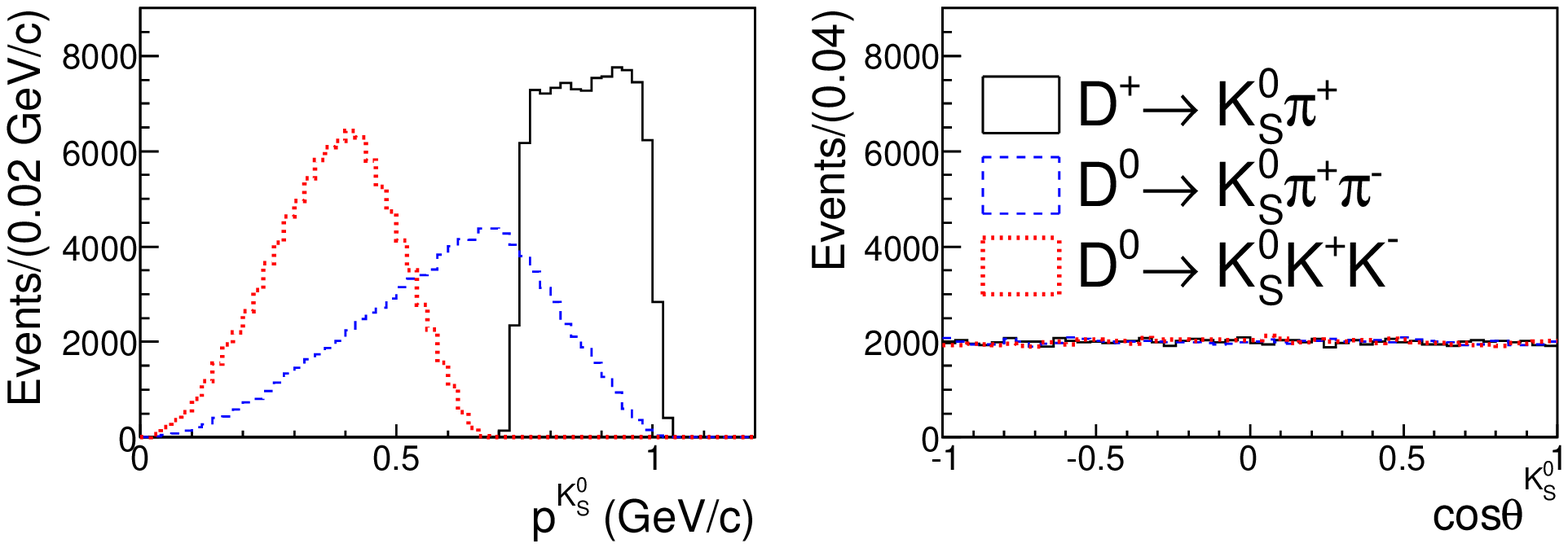}
}
\caption{$K^0_S$ momentum (left column) and angular distributions
  (right column) for different decay modes. The upper two rows are for
  $\sqrt{s}$=10.58 GeV and the $\beta\gamma$=0.425 configuration
  and the bottom row is for $\sqrt{s}$=3770 MeV.}
\label{fig:pna}
\end{figure}

To obtain the expected four-vectors of $K^0_S$ mesons in the final state,
we use {\sc Pythia}~\cite{ref:pythia} and {\sc
 EvtGen}~\cite{ref:evtgen} Monte Carlo codes to simulate generation
and decay of charmed and $B$ mesons produced in $e^+e^-$ collisions.
Two kinematic cases are considered reflecting two 
distinct experimental environments: the first case is for a 
center-of-mass energy $\sqrt{s}$=$10.58$\,GeV and a Lorentz boost
factor of $\beta \gamma$=$0.425$ ($B$ factory), and the second
case is for $\sqrt{s}$=$3770$ MeV with no Lorentz boost (charm factory).

The numerical values of Eq.~(\ref{eq:k0asmII}) are calculated for 
$D^+\rightarrow{K}_S^0 \pi^+$, $D^0\rightarrow{K}_S^0 \pi^+ 
\pi^-$, $D^0\rightarrow{K}_S^0
K^+ K^-$, $B^+\rightarrow{K_S^0} \pi^+$, 
$B^0\rightarrow{K_S^0} \pi^+ \pi^-$, and
$B^0\rightarrow{K_S^0} K^+ K^-$~\cite{ref:conj}, produced in the two 
kinds of 
$e^+e^-$ collisions described above. The choice of the decay channels
is arbitrary, but intended to show a broad range of momenta that 
depend on decay characteristics.  The first four plots in
Fig.~\ref{fig:pna} show the momentum and polar angle distributions of
$K^0_S$ mesons in the laboratory frame for served
final states at $\sqrt{s}$=10.58 GeV and 
$\beta\gamma$=0.425. The distributions in polar angle are
seen to be very similar for $K^0_S$ from charmed and $B$ meson decays,
despite that
the momentum distributions show large differences among
the decay modes, which causes significant differences in the values of
$A_\mathcal{D}$.

\begin{table}
\caption{
\label{table:configuration} 
Two beam pipe and detector configurations selected for the
study described in the text, with $\delta$ and $R$ corresponding to 
the thickness and radius of the
given detector component. There are two
configurations of layers given for Case II.}
\begin{ruledtabular}
\begin{tabular}{cll} 
  & Beam pipe & Detector layers \\
Material & Be  & Si \\
\hline
 & $\delta$=1 mm & $\delta$=300 $\mu$m \\
\raisebox{1.6ex}[0cm][0cm]{Case I} & at $R$=1.5 cm & at $R$=2.0, 4.35, 7.0, 8.8 cm \\
\hline
  & $\delta$=1 mm & $\delta$=50 $\mu$m \\
  & at $R$=1.0 cm & at $R$=1.4, 2.2 cm \\
\raisebox{1.6ex}[0cm][0cm]{Case II}  &  -            & $\delta$=300 $\mu$m \\
  &  -            & at $R$=3.8, 8.0, 11.5, 14.0 cm \\
\end{tabular}     
\end{ruledtabular}
\end{table}

As for the material geometry, we choose two general detector
options, summarized in Table~\ref{table:configuration}, that closely
resemble the existing or planned $B$-meson and 
charm factories. The first option, denoted as 
``Case I''~\cite{ref:belle_beampipe,ref:belle_svd}, reflects the
current charm and $B$-meson-factory experiments. The second option,
denoted as ``Case II'', reflects a proposed super $B$-factory 
experiment~\cite{ref:belle2}.
We apply typical geometrical acceptance 
criteria in calculating $A_{\mathcal{D}}$ for each case. 

\begin{table*}[htbp]
\caption{
\label{table:ad} 
Numerical estimation of $A_\mathcal{D}$ for three
configurations. The values in parentheses are 
only for the beam pipe element. 
}

\begin{ruledtabular}
\begin{tabular}{lccc} 
            & \multicolumn{3}{c}{Configurations} \\[1.0 mm]
            & Case I,  $\sqrt{s}$=10.58 GeV, $\beta \gamma$=0.425 
            & Case I,  $\sqrt{s}$=3770 MeV  
            & Case II, $\sqrt{s}$=10.58 GeV, $\beta \gamma$=0.425 \\
~~Decay Modes & $A_\mathcal{D}$($\times 10^{-4}$)
 & $A_\mathcal{D}$ ($\times 10^{-4}$) 
 & $A_\mathcal{D}$ ($\times 10^{-4}$) \\
\hline
\\[-3.0mm]
$D^+\rightarrow {K}_S^0\pi^+$     &10.8 (~9.0) &15.9 ~(12.0) &~8.8 (~8.5)\\
$D^0\rightarrow {K}_S^0\pi^+\pi^-$&12.9 (11.0) &17.4 ~(14.7) &10.5 (10.4)\\
$D^0\rightarrow {K}_S^0K^+K^-$    &15.1 (12.8) &30.6 ~(27.0) &12.0 (11.8)\\
\hline
\\[-3.0mm]
$B^+ \rightarrow K_S^0 \pi^+      $  & ~6.3 (4.5) & $\dots$  & ~5.2 (4.3)\\
$B^0 \rightarrow K_S^0 \pi^+ \pi^-$  & ~9.1 (7.1) & $\dots$  & ~7.5 (6.7)\\
$B^0 \rightarrow K_S^0  K^+K^-    $  & ~9.5 (7.4) & $\dots$  & ~7.8 (7.0)\\
\end{tabular}     
\end{ruledtabular}
\end{table*}

\begin{figure}[htbp]
\mbox{
\includegraphics[width=0.52\textwidth]{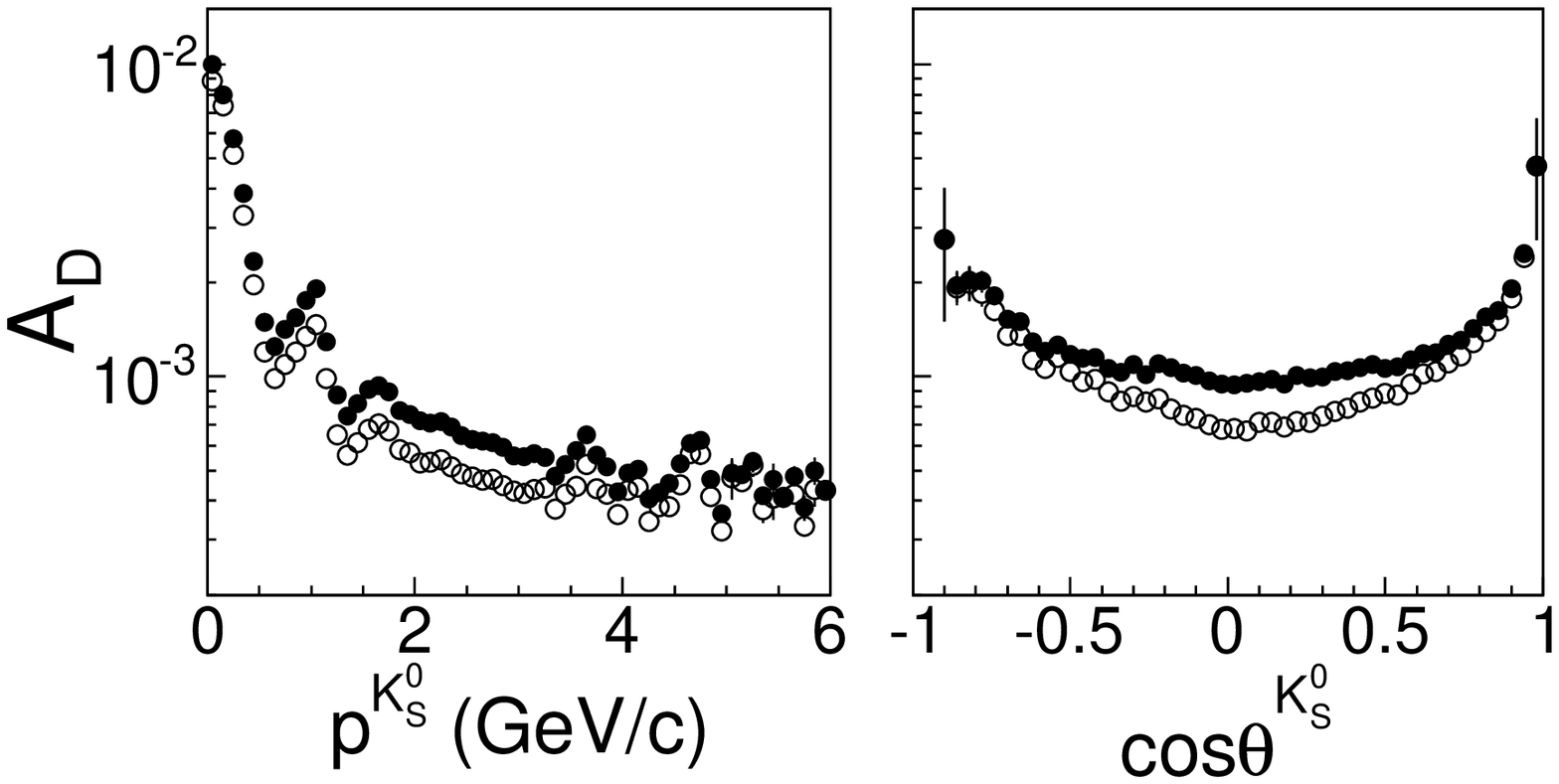}
}
\mbox{
\includegraphics[width=0.52\textwidth]{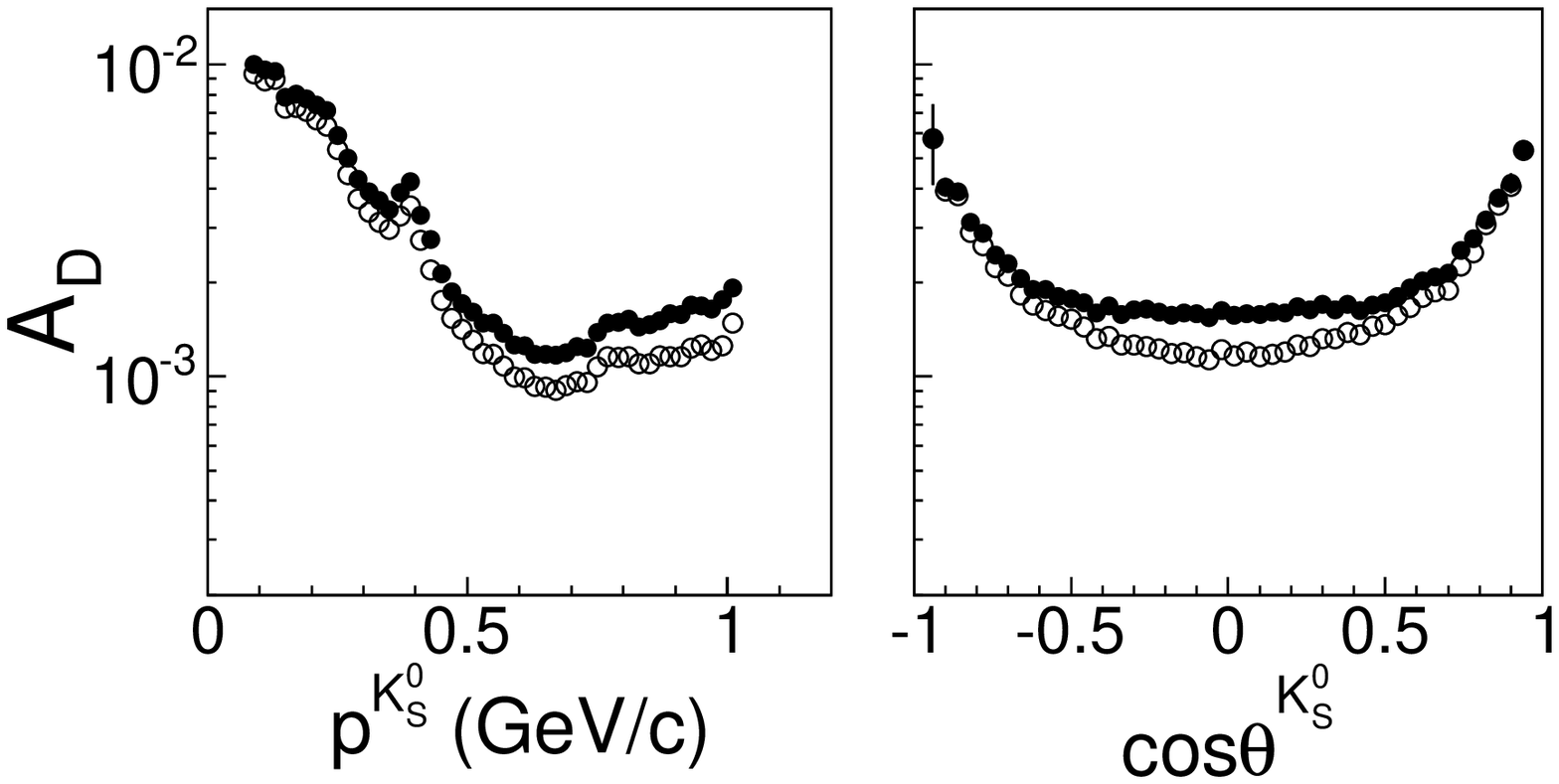}
}
\caption{Distributions of $A_\mathcal{D}$ as a function of $K^0_S$
  momentum (left) and the polar angle (right) for $D^+ \to K^0_S
  \pi^+$ for $\sqrt{s}$=10.58 GeV and $\beta \gamma$=0.425 (top)
  and for $D^0 \to K^0_S \pi^+ \pi^-$ for $\sqrt{s}$=3770 GeV
  configuration (bottom). Case I detector geometry is used in
  both instances.  
}
\label{fig:ad}
\end{figure}

We calculate $A_\mathcal{D}$ for Case I, with
$\sqrt{s}$=10.58 GeV and $\beta\gamma$=0.425, for the decay modes
mentioned previously, and their resultant values are summarized in
Table~\ref{table:ad}. 
We find that $A_\mathcal{D}$ values are $\approx$$10^{-3}$ for 
all the above decay modes, and they are mainly affected by the beam pipe.  We also plot 
the distributions of $A_\mathcal{D}$ as a function of momentum and 
polar angle of $K^0_S$ for Case I.
The upper plots of Fig.~\ref{fig:ad} are the $A_\mathcal{D}$ distributions for 
$D^+ \rightarrow K_S^0 \pi^+$ at $\sqrt{s}$=10.58 GeV 
and $\beta\gamma$=0.425. 
The values of $A_\mathcal{D}$ depend strongly on ${K}^0_S$
momentum distributions as shown in Fig.~\ref{fig:ad} and are larger
for smaller momenta. This can be understood from the
fact that the cross section difference is larger at small momenta
as shown in the upper plot of Fig.~\ref{fig:xsec_pdg}. 
We apply typical experimental selection criteria of
$p^{*}(D^+) >$  2.5 GeV/$c$ and $p_T(\pi^+) >$  0.45 GeV/$c$ in $D^+
\rightarrow K^0_S \pi^+$ decay, where $p^{*}(D^+)$ and
$p_T(\pi^+)$ are the momenta of $D^+$ in the center-of-mass
frame and the transverse momenta of $\pi^+$ in the laboratory
frame, respectively. We find practically no difference in $A_\mathcal{D}$ 
applying these selection criteria.

The major systematic
uncertainty in this calculation is from the assumption 
$\Re (\Delta f)/\Im (\Delta f)$=1. We estimate this effect 
using momentum-dependent values of $\Re (\Delta f)/\Im (\Delta f)$, where
 $\Re (\Delta f)$ is obtained from the best known values 
in Ref.~\cite{ref:martin}
(kaon momenta available up to 2.6 GeV/$c$). The results 
differ from $\Re (\Delta f)/\Im (\Delta f)$=1 
by 6\% when 
limiting the $K^0_S$ momentum range up to 
2.6 GeV/$c$ for charmed-meson decays. 
For $B$ meson decays, the effect is found to be 10\%. 
Because of limited information on 
$\Re(\Delta f)$, we assign systematic uncertainties of
10\% and 20\% for charmed and $B$ meson decays, respectively, for 
the assumption of
$\Re (\Delta f)/\Im (\Delta f)$=1. The systematic effect from 
the assumption that
$\Delta m\approx\frac{1}{2}\Delta \Gamma$ is found to be negligible. 
Systematic effects due to uncertainties in modeling
Eq.~(\ref{eq:brko}) 
are also found to be negligible.
The systematic uncertainties from the measurements for 
$\sigma(K^-\textrm{p})$ and $\sigma(K^+\textrm{p})$ are 0.5\% and 0.9\%, 
respectively. Systematic uncertainties due to the statistical uncertainties on
$\sigma(K^- \textrm{p})$ and $\sigma(K^+ \textrm{p})$ are 
estimated from Monte Carlo,
and found to
be negligible. Other sources include uncertainties on $\Delta m$,
and lifetimes of $K^0_L$ and $K^0_S$, and are also negligible.
There is a contribution from doubly Cabibbo-suppressed decays of  
charmed mesons that is neglected
in the computation of $A_\mathcal{D}$. According to Ref.~\cite{ref:bigi},
we assign a 10\% systematic
uncertainty to the final states with 
a contribution from doubly Cabibbo-suppressed decays.

In the study of the same decay channels of charmed mesons for the
center-of-mass energy in the region of $\psi$(3770), we introduce
no Lorentz boost for the detector geometry described by Case
I. This checks the effect of
different kinematics of $K^0_S$ by comparing the results with those for
$\sqrt{s}$=10.58 GeV and $\beta\gamma$=0.425. The bottom two plots 
in Fig.~\ref{fig:pna} show
the momentum and polar angle distributions of $K^0_S$ mesons for 
$\sqrt{s}$=3770 MeV with no Lorentz boost, showing lower 
$K^0_S$ momentum distributions 
relative to 
those from the configuration with
$\sqrt{s}$=10.58 GeV and $\beta\gamma$=0.425. 
Larger $A_\mathcal{D}$ values are consequently expected, which is 
consistent with the
calculations shown in the third column of Table~\ref{table:ad}.  
The bottom of Fig.~\ref{fig:ad} shows the distributions of $A_\mathcal{D}$ 
as a function of momentum and 
polar angle of $K^0_S$ for $D^0\rightarrow K_S^0 \pi^+\pi^-$
at $\sqrt{s}$=3770 MeV and $\beta\gamma$=0.
We find that the $A_\mathcal{D}$ values are in general larger than given 
in the second column of Table~\ref{table:ad}.
Again, this reflects the $K^0_S$ momentum distribution shown in the bottom plot 
in Fig.~\ref{fig:pna},
which peaks in the phase space 
region with the largest $\Delta \sigma(\bar{K}^0 N)$. 
Here, the systematic uncertainty from the assumption 
that $\Re(\Delta f)/\Im(\Delta f)=1$ 
is found to be 30\%, and other sources are negligible.

As a final benchmark, we also evaluate $A_\mathcal{D}$ for the Case
II configuration with $\sqrt{s}$=10.58 GeV and $\beta
\gamma$=0.425. The results of estimating of $A_\mathcal{D}$ are
listed in the last column of Table~\ref{table:ad}. This
configuration checks the effect of different geometry for detector 
material by comparing results from Case I with the same 
kinematics. We find that the contribution of the first two
thin layers of Si sensors is negligible. Furthermore,
the contribution of the outer Si sensors is also smaller than 
that of the Si sensors in Case I as their distances from the production 
point of neutral kaons are longer. This results in smaller dilution 
than for Case I. Systematic sources and effects are
similar to those of due to Case I.
 
As shown above, the dilution effect in the calculation of $A_\mathcal{D}$ is 
most sensitive to the momentum of $K^0_S$,
and mainly due to the beam pipe contribution.
Hence, the dilution effect in very high energy experiments in the LHC 
environment 
can be smaller than the impact in experiments considered in this paper.

In summary, we estimate the dilution effect in the 
measurement of $CP$ asymmetry 
caused by the difference in nuclear interactions of 
$K^0$ and $\bar{K}^0$ in $e^+e^-$ 
collisions for several typical experimental configurations. 
We find that the effect 
can be as large as 0.3\% in decays
involving low-momentum neutral kaons.
The estimated systematic uncertainties on the calculated $A_\mathcal{D}$ 
range in (15$\approx$30)\% depending on $K^0_S$ momentum. 
We suggest that forthcoming 
high-sensitivity measurements of $CP$ asymmetry 
involving neutral kaons in the final state 
should take into account the impact on 
the difference in $K^0$ and $\bar{K}^0$ strong interactions 
($A_\mathcal{D}$).

The authors would like to thank Simon I. Eidelman for his help in the
preparation of the paper.
E. Won 
acknowledges support by NRF Grant No. 2011-0027652 and B. R. Ko
acknowledges support by NRF Grant No. 2011-0025750.


\begin{thebibliography}{99}
\bibitem{ref:conj} Throughout this paper, the charge-conjugate state is
  implied unless stated otherwise.
\end{thebibliography}
\end{document}